# Dynamic Mode Decomposition for data-driven analysis and reduced-order modelling of E×B plasmas: I. Extraction of spatiotemporally coherent patterns

F. Faraji*[1], M. Reza*, A. Knoll*, J. N. Kutz**

* Plasma Propulsion Laboratory, Department of Aeronautics, Imperial College London, London, United Kingdom

** Department of Applied Mathematics and Electrical and Computer Engineering, University of Washington, Seattle, United States

**Abstract**: The advent of data-driven/machine-learning based methods and the increase in data available from high-fidelity simulations and experiments has opened new pathways toward realizing reduced-order models for plasma systems that can aid in explaining the complex, multi-dimensional phenomena and enable forecasting and prediction of the systems' behavior. In this two-part article, we evaluate the utility and the generalizability of the Dynamic Mode Decomposition (DMD) algorithm for data-driven analysis and reduced-order modelling of plasma dynamics in cross-field E×B configurations. The DMD algorithm is an interpretable data-driven method that finds a best-fit linear model describing the time evolution of spatiotemporally coherent structures (patterns) in data. We have applied the DMD to extensive high-fidelity datasets generated using a particle-in-cell (PIC) code based on a cost-efficient reduced-order PIC scheme. In this part, we first provide an overview of the concept of DMD and its underpinning Proper Orthogonal and Singular Value Decomposition methods. Two of the main DMD variants are next introduced. We then present and discuss the results of the DMD application in terms of the identification and extraction of the dominant spatiotemporal modes from high-fidelity data over a range of simulation conditions. We demonstrate that the DMD variant based on variable projection optimization (OPT-DMD) outperforms the basic DMD method in identification of the modes underlying the data, leading to notably more reliable reconstruction of the ground-truth. Furthermore, we show in multiple test cases that the discrete frequency spectrum of OPT-DMD-extracted modes is consistent with the temporal spectrum from the Fast Fourier Transform of the data. This observation implies that the OPT-DMD augments the conventional spectral analyses by being able to uniquely reveal the spatial structure of the dominant modes in the frequency spectra, thus, yielding more accessible, comprehensive information on the spatiotemporal characteristics of the plasma phenomena.

## Section 1: Introduction

With increased computational power and abundant generation of data over the last decades, the traditional paradigms of simulation and analysis across the sciences have undergone a fundamental transformation. Data-driven techniques have emerged as a cornerstone of contemporary research, offering novel ways to either analyze data and extract insights or to develop predictive models from large and complex datasets. Data-driven techniques depart from the conventional reliance on predefined models or theoretical assumptions. Rather, these techniques process data to uncover underlying patterns, information, relationships, and trends that might otherwise remain obscure. These approaches have proven particularly powerful where the traditional approaches fall short, such as complex systems with high-dimensional data spaces, and situations where first-principles generalizable equations are challenging to derive.

Modeling and understanding the physics of plasmas is an area that can markedly benefit from data-driven approaches. The multiscale nature of the underlying physics and the intricate interplay of various phenomena across multiple dimensions render the traditional high-fidelity plasma simulations computationally intensive and/or prohibitive due to their high dimensionality. Reduced-order models (ROMs) seek to alleviate this computational burden/intractability by approximating a system's behavior using a lower-dimensional representation while preserving its essential dynamics. This reduced-order representation enables faster simulations, real-time control, and optimization of complex systems. The reduced-order representation also helps our understanding of the complex interactions between various processes in the plasma, such as instabilities, by distilling the high dimensional dynamics and providing a more interpretable and lower dimensional description.

Dynamic Mode Decomposition (DMD) is a data-driven technique that was introduced by Schmid [1] in fluid dynamics to decompose a flow field into series of coherent spatiotemporal modes with exponential time dynamics. Due to its simplicity of implementation and interpretability, the DMD method has ever since gained remarkable popularity and has been used in various applications, particularly in the field of fluid dynamics [2]-[4].

---
[1] **Corresponding Author** (f.faraji20@imperial.ac.uk)

The recent development of the optimized DMD (OPT-DMD) [5] and the bagging optimized DMD (BOP-DMD) [6] has significantly changed the potential of DMD to perform critical modeling tasks due to the robustness of these new algorithms to noise. Importantly, OPT-DMD can be used to constrain best-fit linear models to be stable by construction (eigenvalues are constrained to the left-half plane), purely oscillatory (eigenvalues are constrained to the imaginary axis), or complex conjugate pairs (solutions are real). Thus, OPT-DMD can be used to address two main challenges of plasma modeling: (a) extraction of robust and stable spatiotemporal coherent structures, and (b) reduced order modeling. Regarding (a), OPT-DMD achieves this by isolating the dynamic modes and their associated spatial structures that characterize the system's evolution. With respect to (b), by retaining a subset of the extracted modes that represent the most salient behaviors of the system, the OPT-DMD effectively constructs a low-dimensional model that can adequately capture the system's dynamics.

It is noted that the DMD method has been also applied to plasma physics [7]-[9], most notably by Taylor et al [10] and Kaptanoglu et al [11] in a z-pinch experimental configuration. These two latter studies provided interpretable insights into the dominant patterns of activity of the z-pinch experiment. However, the OPT-DMD algorithm had not yet been developed at the time of those works, which precluded these studies from generating stable reduced order models and/or accurate representations of the power spectral densities. These considerations underline the novelty and timeliness of the present effort.

We also emphasize that a DMD model, in general, is known to be particularly useful for short-to-medium term predictions due to its limitations and underlying assumptions that include model linearity and the dominance of a fixed set of DMD modes [12]. In many real-world systems, linearity holds reasonably well over short to medium time scales. However, as the prediction horizon extends, non-linear effects and interactions can become more significant, leading to deviations from the linear assumption. This can limit the accuracy of the DMD for longer-term predictions. In any case, we will see in part II that, for plasma systems with a quasi-periodic behavior at the steady-state, the error in the forecast will remain "bounded" even for extended forecast time horizons.

With regard to the DMD's assumption of the dominance of a fixed set of modes, as this method models a system's behavior based on historical data, if the dynamics of the system changes, the existing DMD predictions might not accurately represent the new behavior. In this regard, over short-to-medium term horizons, a certain number of modes can indeed dominate the system's behavior, making DMD predictions reliable. However, in a system with non-steady dynamics, different sets of modes may be dominant over different time intervals which represents the transient phenomena in the system. Even in these circumstances, a rather short-term prediction window could be sufficient in certain applications where there is the possibility to periodically receive real-time data from the physical system. In such cases, the DMD model can be adapted to reflect the change in the system dynamics. This can be achieved by recalculating the modes using the received data and, thus, updating the model as new observations become available. Thus, the prediction horizon of a DMD model in practice only needs to be longer than the time it takes to collect the data and re-fit the model. This enables the application of the DMD, despite its simplicity and limitations, toward control of systems' evolution and the development of digital system models and twins to digitally represent/mimic the behavior of physical systems.

As a result, when assessing the accuracy of a DMD model, it is important to take into account the reduction in the model order and the complexity, ensuring a balanced evaluation of the model's capabilities in relation to the computational resources it requires. With this in mind, we demonstrate in this two-part article that the DMD approach can be, in general, a powerful tool for the study and modeling of plasmas.

We assess in this work the application of the DMD method to partially magnetized plasmas immersed in a perpendicular arrangement of the electric and magnetic fields, which are typically referred to as cross-field plasmas. Using data from high-fidelity particle-in-cell (PIC) simulations in several conditions, we assess the capability of the DMD to derive the coherent spatiotemporal modes in the data as well as to develop reduced-order models. We have chosen E×B plasma configurations for this study since they are characterized by a variety of instabilities across a wide range of temporal and spatial scales, and their dynamics are dominated by complex interactions between co-existing instabilities. This intriguing physics has in fact made the E×B discharges the subject of numerous scientific research [13]. Moreover, the E×B plasma configurations are additionally of great industrial interest as evidenced by their presence across the low-temperature plasma technologies including Hall thrusters for in-space plasma propulsion and magnetrons for material processing.

In part I of the paper, in an E×B plasma discharge representative of a radial-azimuthal cross-section of a Hall thruster, we focus on demonstrating the efficacy of the DMD in a range of plasma conditions toward the extraction of the spatiotemporal coherent structures in order to distinguish and isolate the individual instability modes. In part II, we will assess the extent of applicability of the DMD for the development of reduced order models and evaluate the forecasting capabilities of these DMD-derived ROMs.

**Section 2: Overview of the DMD method**

The DMD is a technique to decompose the dynamics of a system into a combination of various spatiotemporal modes with arbitrary spatial bases and linear time dynamics. Therefore, unlike Fourier transform which assumes purely harmonic basis functions, DMD modes are specific to each system and are driven directly from the data. The time dynamics of the DMD modes are constrained to be linear, which means that each spatial mode is associated with a complex frequency representing an oscillatory behavior possibly with a growth or damping rate.

DMD can be thought of as an alternative dimensionality reduction technique to Singular Value Decomposition (SVD) or Proper Orthogonal Decomposition (POD). Indeed, the SVD/POD modes are orthogonal and, therefore, provide an optimum set of bases for representing the data [12]. However, the SVD/POD modes do not contain any temporal evolution information. In contrast, the DMD provides dynamical information about the system at the expense of the fact that the modes are no longer orthogonal because they are constrained by the linearity of the time dynamics. Below, we provide a brief overview of the SVD/POD first, followed by an introduction to the algorithm of basic DMD [14] in Section 2.2. There are several promising DMD variants as well, including optimized DMD [5], which we also mentioned in Section 1, the multi-resolution DMD (mrDMD) [15], high-order DMD (HODMD) [16], and parametric DMD [17]. In Section 2.3, we briefly introduce the optimized DMD variant based on variable projection. Our focus within this two-part article will be on the application of this DMD variant to our plasma system data.

**2.1. Proper Orthogonal Decomposition (POD) and Singular Value Decomposition (SVD)**

POD, also known as Principal Component Analysis (PCA) in some contexts, is a technique used for dimensionality reduction, data compression and noise reduction. It aims to capture the dominant patterns or modes present in a dataset. In POD, we start with a set of data, often represented as snapshots or measurements taken over time or space. The goal is to find a reduced set of basis functions, also called modes, that can accurately approximate the original data. These modes are determined based on the statistical properties and the correlations of the data, and they are orthogonal (perpendicular) to each other, which helps in capturing effectively the most important variations in the data. POD is closely related to SVD [12], which is a matrix factorization technique that enables computing the basis modes.

SVD decomposes a data matrix $X \in C^{n \times m}$ (whose columns can be measurements in space or time) into three matrices, $U \in C^{n \times m}$, $\Sigma \in R^{m \times m}$, and $V^* \in C^{m \times m}$, hence,

$$X = U\Sigma V^*. \tag{Eq. 1}$$

$n$ denotes the dimensionality of the data, and $m$ represents the number of measurements or snapshots. In addition, in Eq. 1, the superscript * indicates conjugate transpose of matrix $V$. The columns of matrix $U$ are the left singular vectors, which represent the SVD/POD modes. These vectors are orthonormal and form a basis for the column space of $X$. These modes capture the correlation between the columns of matrix X (different snapshots), and they are arranged in a hierarchical order from left to right based on the extent of correlation they represent. Matrix $V$ contains the right singular vectors. Similar to $U$, the vectors in $V$ are orthonormal as well and provide a basis for the row space of $X$. The columns of $V$ capture the correlation between rows of X and, hence, in a time series data, they represent the temporal pattern. $\Sigma$ is a diagonal matrix with non-negative entries $\sigma_i$, $i = 1,2,..m$ along its diagonal. These $\sigma_i$ entries are the singular values associated with the SVD/POD modes. The singular values are arranged in descending order and represent the "strength" or "importance" of the corresponding singular vectors in matrices $U$ and $V$.

In practice, it is not necessary to retain all the SVD modes (i.e., full $U$, $V$ and $\Sigma$ matrices) to be able to represent the data. In many cases, the data matrix can be well approximated by a lower-rank matrix because the dominant patterns and significant information in the data are captured by a subset of these modes. The dominant patterns are captured by the larger singular values and their corresponding singular vectors in SVD. Therefore, we can truncate the SVD by considering only the most significant singular values and their corresponding columns in matrices $U$ and $V$. This reduces the dimensionality of the data while retaining its important features. Besides, real-world data often contains noise or insignificant information represented by smaller singular values. By truncating these smaller singular values and their associated modes, we can effectively remove the noise and irrelevant information, focusing on the more significant patterns.

The low-rank truncation and determining how many ranks to retain is performed by looking at the distribution of the singular values ($\sigma_i$). The ratio given by Eq. 2 determines the proportion of the total variance in the data that is captured by the first k singular values and their corresponding modes.

$$K = \frac{\sum_{i=1}^{k} \sigma_i}{\sum_{i=1}^{m} \sigma_i}. \tag{Eq. 2}$$

Typically, we aim to retain around 90-95% of the total variance to capture the most important features. The rank-$r$ approximation of the data ($\tilde{X}_r$) is obtained by keeping the first $r$ columns of $U$ ($u_i$) and $V$ ($v_i^*$) and first $r$ singular values ($\sigma_i$) as per Eq. 3.

$$\tilde{X}_r = \sum_{i=1}^{r} \sigma_i u_i v_i^* = \tilde{U}\tilde{\Sigma}\tilde{V}^*. \tag{Eq. 3}$$

In Eq.3, $\tilde{U} \in C^{n\times r}$, $\tilde{\Sigma} \in R^{r\times r}$ and $\tilde{V} \in C^{m\times r}$.

## 2.2. The basic DMD algorithm

The central idea of DMD is to provide a best-fit linear dynamics approximation of the spatiotemporal data from either a linear or non-linear system. The basic algorithm of the DMD is outlined in the following.

The collected time-series data representing the time evolution of the system (which can come from either an experiment or a high-fidelity simulation) are rearranged into data matrices $X$ and $X' \in R^{n\times m-1}$

$$X = \begin{bmatrix} | & | & & | & & | \\ x_1 & x_2 & \ldots & x_k & \ldots & x_{m-1} \\ | & | & & | & & | \end{bmatrix}, \quad X' = \begin{bmatrix} | & | & & | & & | \\ x_2 & x_3 & \ldots & x_{k+1} & \ldots & x_m \\ | & | & & | & & | \end{bmatrix}.$$

Each column of $X$ ($x_k$) represents an n-dimensional state vector at time step $t_k$ where $k = 1, 2, \ldots m-1$, and the corresponding column in $X'$ is the system's state vector in the subsequent time step $t_{k+1}$. In our context, $x_k$'s are the snapshots of the plasma flow field at each time step of the simulation rearranged into a column vector. The DMD algorithm finds the optimal linear transformation/operator ($A$) that approximate simultaneous advancement of all data one time-step forward such that

$$X' \approx AX, \tag{Eq. 4}$$

with $A$ defined as,

$$A = \arg\min_{A} \|x_{k+1} - Ax_k\|_2. \tag{Eq. 5}$$

The solution of this minimization problem is

$$A = X'X^\dagger = X' V\Sigma^{-1} U^*, \tag{Eq. 6}$$

where $X^\dagger$ is Moore-Penrose pseudo-inverse of $X$ obtained using the SVD of matrix $X$.

In any case, typically, the snapshots are high dimensional such that $n \gg m$, which means that matrix $A \in R^{n\times n}$ can be prohibitively large to compute and store. As such, the DMD relies on SVD/POD to project matrix $A$ onto a reduced-dimensional subspace defined by $r$ POD modes and find a reduced rank approximation of the full matrix $A$ denoted by $\tilde{A} \in R^{r\times r}$. Therefore, the next step in calculating the DMD is to take the SVD of data matrix $X$ and project matrix $A$ onto the rank-r truncated SVD modes $\tilde{U}$ to obtain $\tilde{A}$

$$\tilde{A} = \tilde{U}^* A \tilde{U} = \tilde{U}^* X' \tilde{V} \tilde{\Sigma}^{-1}. \tag{Eq. 7}$$

Having obtained $\tilde{A}$, we compute the eigendecomposition of matrix $\tilde{A}$ to obtain the DMD spatial modes and their dynamics

$$\tilde{A}V = \Lambda V. \tag{Eq. 8}$$

In Eq. 8, $\Lambda$ is a diagonal matrix containing the eigenvalues ($\lambda_i$, i=1, 2, ..., r) of $\tilde{A}$, which coincide with the leading eigenvalues of the full matrix $A$. The columns of matrix $V$ are the eigenvectors of $\tilde{A}$, which can be expanded to reconstruct the eigenvectors of $A$ (represented by columns of $\Phi$ ($\phi_i$, i=1, 2, ..., r)) using the relation in Eq. 9.

$$\Phi = X'V\Sigma^{-1}V. \tag{Eq. 9}$$

Each eigenvalue-eigenvector pair ($\phi_i$ and $\lambda_i$) represents a distinct DMD spatiotemporal mode, where $\phi_i$ characterizes the spatial behavior and $\lambda_i$ describes temporal evolution of the mode. The reconstructed time series of state vectors from the DMD modes is obtained through their linear combination

$$x_{k+1} \approx \sum_{i=1}^{r} b_i \phi_i \lambda_i^k = \Phi B \Lambda^k, \qquad k = 1, 2, \ldots \tag{Eq. 10}$$

where, matrix B is a diagonal matrix containing $b_i's$, which are the initial amplitude of the i-th DMD mode. The modes initial amplitudes are obtained by projecting the data at $t = 0$ ($x_1$) onto DMD modes.

$$B = \Phi^\dagger x_1 \tag{Eq. 11}$$

Eq. 10 represents the discrete form of the dynamics describing the evolution of the system over discrete time intervals $\Delta t$ corresponding to the time step at which the measurements are taken. This equation can also be written in the continuous dynamics form as in Eq. 12.

$$x(t) \approx \sum_{i=1}^{r} b_i \phi_i \exp(\omega_i t) = \Phi B \exp(\Omega t). \tag{Eq. 12}$$

$\Omega$ is a diagonal matrix that contains the continuous-time eigenvalues ($\omega_i$, $i = 1, 2, \ldots, r$) on its diagonal. $\omega_i$ and $\lambda_i$ are related through the relationship below

$$\omega_i = \ln(\lambda_i)/\Delta t. \tag{Eq. 13}$$

Eq. 12 and Eq. 13 can be used for forecasting the system's behavior in future times.

**2.3. Optimized DMD method based on variable projection**

One of the major limitations of the basic DMD algorithm, which is referred to as the "Exact" DMD in this paper, is its susceptibility to bias errors from noisy data, which can lead to poor model fits and unstable forecasting capabilities. This sensitivity to noise arises from the fact that the DMD considers the relationship between the consecutive snapshot pairs in the dataset. Therefore, the presence of noise can mis-inform the pairwise relation between the snapshots, which can disrupt accurate identification of underlying patterns and modes in the data. Moreover, the identified modes can involve notable growth or damping rates resulting in the model predictions to either vanish to zero or go to infinity over longer terms. In this regard, noting that most real-world data contain certain noise levels, the standard Exact DMD algorithm leads to unstable and inaccurate models, which are incapable of reproducing data over practically desirable long durations.

The optimized DMD algorithm based on variable projection [5] represents a remarkable advancement from the Exact DMD method. This so-called "OPT-DMD" particularly circumvents the limitation of the Exact DMD with respect to noise-induced bias, making the DMD method more stable, robust, and predictive. OPT-DMD mitigates the sensitivity to noise by accounting for the relationship among all snapshots collectively rather than relying exclusively on pairwise connections. This algorithm allows us to minimize the real part of the eigenvalues, thus, constraining the modes to have purely oscillatory dynamics and as a result stabilizing the model which will be in turn capable of excellent reconstruction of the original (ground-truth) dataset. Additionally, OPT-DMD can function with non-equispaced snapshots in time, which is another notable improvement compared to the Exact DMD.

OPT-DMD relies on variable projection technique to efficiently solve the following nonlinear least-square optimization problem to find $\Omega$ and $\Phi_B = \Phi B$ [5]

$$\min_{\Omega, \Phi_B} \|X - \Phi_B \exp(\Omega t)\|_2. \tag{Eq. 14}$$

The mathematical details of the OPT-DMD method and the approach to solve the optimization problem given by Eq. 14 are beyond the scope of this article. Interested readers are encouraged to refer to the original article on OPT-DMD [5].

To apply the OPT-DMD algorithm to the plasma system data in this two-part article, we have used the open-source MATLAB code of Ref. [5] available at [18].

## Section 3: Overview of the radial-azimuthal cross-field discharge test cases

The test cases used for the reported investigations in this paper are identical in simulation setup and parameter range to those in our previous publication [19] related, in part, to the study of the effects of plasma number density on the physics of the instabilities and electron transport in a cross-field discharge representative of a Hall thruster's radial-azimuthal cross-section.

To provide a brief overview, the reduced-order quasi-2D simulations [20]-[22] of the radial-azimuthal plasma configuration were carried out in a domain corresponding to a 2D $x-z$ Cartesian plane. For these simulations, the $x$-coordinate is along the radial direction, and the $z$-coordinate represents the azimuthal direction. The y-axis is along the axial direction. The domain is 1.28-cm long along both the radial and azimuthal directions. Other details of the simulations' setup, such as the initial conditions, particle loading, and potential and particle boundary conditions are provided in Ref. [19][24].

The domain decomposition associated with the reduced-order PIC scheme [20] has been carried out using 50 vertical regions ($M$) and 50 horizontal regions ($N$). This translates into quasi-2D simulations with a five-fold reduction in the computational cost compared to the corresponding conventional full-2D PIC simulations [23].

The radial-azimuthal test-case simulations feature a temporally invariant cosine-shaped ionization source along the radial direction with the peak value of $S$ to compensate for the flux of particles lost to the walls [19][24]. The simulations also include a constant axial electric field ($E$) and a constant radial magnetic field ($B$) [19].

The test cases in this part I paper correspond to simulations with different values of $S$, which are equivalent to various plasma number densities at the system's quasi-steady state. The sets of physical parameters associated with these test cases are summarized in Table 1.

| Test case | S value | E value | B value |
|---|---|---|---|
| 1 (Baseline) | $S_0 = 8.9 \times 10^{22} \ m^{-3} s^{-1}$ | $E_0 = 10,000 \ Vm^{-1}$ | $B_0 = 0.02 \ T$ |
| 2 | $1/32 \ S_0$ | $E_0$ | $B_0$ |
| 3 | $1/16 \ S_0$ | $E_0$ | $B_0$ |
| 4 | $1/8 \ S_0$ | $E_0$ | $B_0$ |
| 5 | $1/4 \ S_0$ | $E_0$ | $B_0$ |
| 6 | $1/2 \ S_0$ | $E_0$ | $B_0$ |
| 7 | $3 S_0$ | $E_0$ | $B_0$ |
| 8 | $6 S_0$ | $E_0$ | $B_0$ |

Table 1: Summary of the values of the physical parameters for various radial-azimuthal test-case simulations; the baseline case has the $S$ value equal to that of the radial-azimuthal benchmark reported in Ref. [24].

In each test case, the total number of plasma properties' snapshots is 2000 corresponding to 30 $\mu s$ of simulation time and each snapshot represents an average over $1.5 \times 10^{-2} \ \mu s$ (1000 timesteps). The first 5.25 $\mu s$ (350 snapshots) of the simulations, which captures the initial transient of the system before reaching a quasi-steady state, is excluded from the dataset. This leaves us with a 1650-snapshot training dataset for each test case.

## Section 4: Results and discussion

This section is structured as follows: we first present in Section 4.1 the results of the application of SVD to our data, in part to assess the extent to which the data has a low-rank representation. We discuss the influence of the truncation rank ($r$) and the training data ratio on the SVD. In Section 4.2, we compare the application of the Exact DMD and the OPT-DMD algorithms, highlighting the capability of the OPT-DMD to reliably reconstruct the ground-truth data over time. We will also evaluate the effect of the DMD truncation rank on the accuracy of the OPT-DMD-reconstructed data. For the investigations in Sections 4.1 and 4.2, we have focused on the data of the baseline radial-azimuthal test case.

In Section 4.3, we present a novel application of the pattern extraction using OPT-DMD toward simultaneous analysis of the spatial-temporal characteristics of the plasma phenomena, such as instabilities and fluctuations, in complementarity or as an alternative to the conventional Fast Fourier Transform (FFT) analyses. To the best of the author's knowledge, our work is the first effort that demonstrates the utility of DMD for this purpose. The results presented in this section are obtained for all test cases mentioned in Table 1. Following on the discussions in Section 4.3, and using the data of the baseline test case, Sections 4.4 and 4.5 assess, respectively, the sensitivity of the OPT-DMD for spatiotemporal characterization to the truncation rank and the general rank inflation issue of the DMD method.

**4.1. Singular Value Decomposition for dimensionality reduction of high-fidelity simulations' data**

We begin the discussions by showing the SVD results to gain insights into the structure and dimensionality of our data as well as to investigate whether the data can be represented with low-enough number of ranks. For these analyses, we have focused on the baseline test case simulation and have taken the axial electron current density ($J_{ey}$) data as a representative plasma property since the influence of the main instabilities and dynamical processes are evidently reflected in the $J_{ey}$ signal.

Figure 1(a) and (b), respectively, present the normalized distribution of the singular values ($\sigma_r / \sum_{i=1}^{m} \sigma_i$) and their normalized cumulative sum ($\sum_{i=1}^{r} \sigma_i / \sum_{i=1}^{m} \sigma_i$) in the SVD of $J_{ey}$ vs the number of rank. From these plots, we can identify the significance of each mode and how much of the variance within the data is captured by keeping the first $r$ ranks. The maximum number of ranks is equal to the number snapshots ($m$), which is equal to 1650 for the radial-azimuthal simulations' data. Figure 1 illustrates that the entire ranks (or SVD modes) are, however, not necessary to faithfully reconstruct the original signal. In fact, the "knee" point in the profiles of the singular values indicates that the major part of the useful information within the data are represented by the first 400-500 modes.

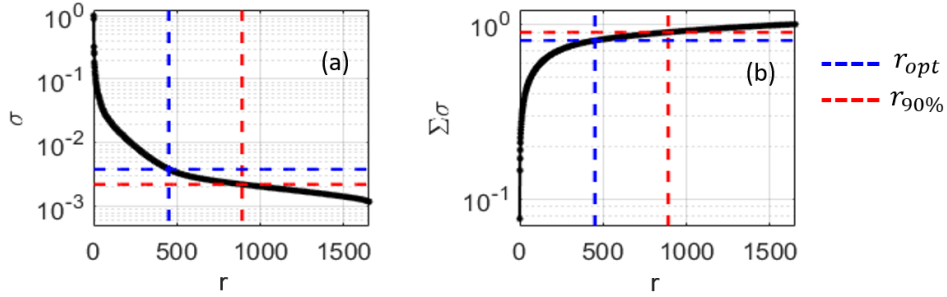

Figure 1: The distribution of (a) normalized singular values and (b) normalized cumulative sum of first $r$ singular values in the SVD of axial electron current density ($J_{ey}$) data.

To quantitatively determine the minimum SVD rank necessary for a high fidelity reconstruction of the original data, there can be two possible approaches: (1) we can consider the minimum $r$ value ($r_{90\%}$) as the fewest number of modes that carry at least 90% of the total variance in the data, i.e., $\frac{\sum_{i=1}^{r} \sigma_i}{\sum_{i=1}^{m} \sigma_i} \geq 0.9$. (2) Ref. [25] has introduced an "Optimal Hard Threshold", which determines the noise floor of the data and that corresponds to a number of ranks ($r_{opt}$) beyond which the modes represent the noise. The number of necessary ranks determined by either of these criteria are shown as dashed lines in Figure 1. The Optimal Hard Threshold (dashed blue line) suggests $r = 450$ for the low-rank approximation of the data matrix. This is much smaller than $r_{90\%}$ (dashed red line), which gives a required rank of about 900. Regardless, it will become clear later in this section that, even with a much lower number of ranks than those given by either criterion, we can still reconstruct the original data with sufficiently high fidelity.

At this point, it is of interest to assess the effectiveness of the SVD bases in representing unseen ("test") data, i.e., data which has not been included in the "training" dataset for which the SVD is calculated. In the context of SVD, testing implies whether the projection of the unseen data onto the SVD modes calculated using the training dataset can provide an accurate low-rank representation of those unseen data.

For this purpose, we divide the entire data into a training and a test set with three different proportions of $R_t =$ 0.2, 0.5, and 0.8. $R_t$ denotes the ratio of the length of the training set to the total dataset. The SVD is calculated on the training dataset, and the reconstruction of the training data is obtained from the rank-$r$ truncated SVD as given by Eq. 3. The average reconstruction loss ($\mathcal{L}$) with respect to the original data for different number of ranks is then calculated using Eq. 15. In Eq. 15, $\|S_i\|_F$ denotes the Frobenius norm of the i-th data snapshot, $N_i$ and $N_j$ are the number of grid points along the horizontal and vertical directions, respectively, and $s_{ij}$ represents the value of the $J_{ey}$ at each node $ij$.

$$\mathcal{L} = \frac{1}{m}\sum_{i=1}^{m}\frac{\left\|S_i^{true} - S_i^{SVD}\right\|_F}{\left\|S_i^{true}\right\|_F}, \quad \|S\|_F \equiv \sqrt{\sum_{i=1}^{N_i}\sum_{j=1}^{N_j}|s_{ij}|^2}, \quad \text{(Eq. 15)}$$

$$x_{k,r} = U_r U_r^T x_k \quad \text{(Eq. 16)}$$

The test data snapshots ($x_k$) are next projected on the SVD modes ($U_r$) according to Eq. 16, which yields rank-$r$ reconstruction of the test data. Similar to the training set, the average loss of reconstruction of the test data with various SVD modes is computed using Eq. 15.

Figure 2 presents the results in terms of the variation against the number of ranks of the losses in the reconstructed training (plot (a)) and test (plot (b)) datasets. It is evident from Figure 2 that the loss in data reconstruction is smaller when using larger dataset to calculate the SVD. This is because, with larger $R_t$ values, more statistics of the signal are made available to the algorithm which results in the identification of more representative and accurate modes to describe the data.

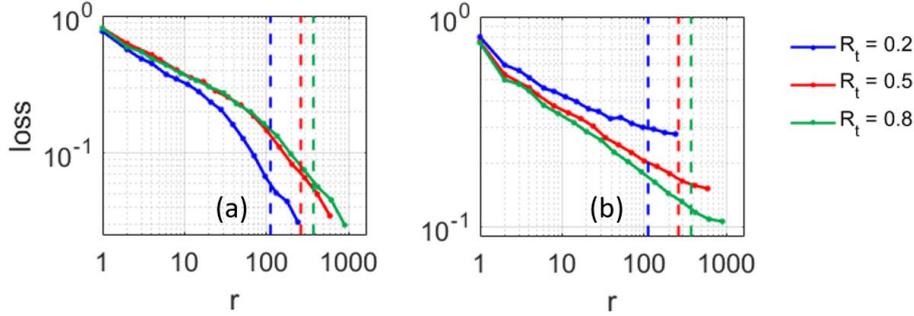

Figure 2: Variation vs rank number ($r$) of the loss factor (Eq. 15) between the ground-truth simulation data (axial electron current density $J_{ey}$) of the radial-azimuthal baseline test case and the reconstructed data from the SVD. Various curves correspond to different ratios of the training to test data ($R_t$). Plot (a) represents the variation in the loss on the training data, whereas plot (b) shows the loss factor variation over the test data.

To further demonstrate the efficacy of the SVD in approximating data in a reduced-order coordinate, we present in Figure 3 a comparison of the time series of spatially averaged values of the reconstructed $J_{ey}$ with various number of ranks and $R_t$ ratios against the original signals from the PIC simulations. The evolutions of the loss between individual original and reconstructed snapshots over time are also superimposed.

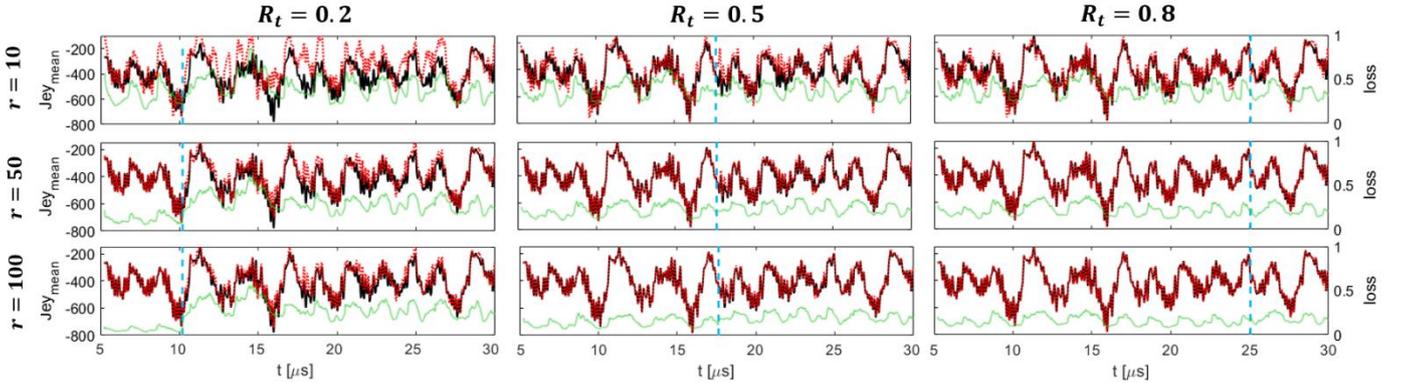

Figure 3: Time evolution of the spatially averaged axial electron current density ($J_{ey}$) from the quasi-2D PIC simulation of the radial-azimuthal test case with the baseline conditions (solid black lines) and the reconstructed signal from the SVD (dotted red lines) using different number of ranks and training-to-test data ratios. Superimposed on each plot is the time evolution of the loss factor (Eq. 15) for the corresponding ($r$, $R_t$) value pair. Dashed blue lines indicate the training end time.

It is impressive that, when a large enough dataset is used to calculate the SVD ($R_t = 0.5$ and $0.8$), even 10 ranks can capture the main information content so that the reconstructed $J_{ey}$ signal can follow closely the original signal. With 50 ranks, the reconstruction is almost in perfect agreement with the original signal for $R_t$ values of 0.5 and 0.8. Beyond 50 ranks (as observed for instance with 100 ranks), any further enhancement in accuracy becomes negligible. The reconstruction of the test data in the case of $R_t = 0.2$ is rather poor and includes noise. This is because the statistical information of 20% of the signal has not been sufficient for the SVD modes to be representative of the entire signal.

In Figure 4, an example of the reconstructed 2D snapshots with different ranks and $R_t$ ratios are provided. These figures convey similar information as those inferred from Figure 2 in a more visual sense. The reconstructed snapshots with 50 ranks capture all features of the original snapshot, which is denoted as "True". We further observe that the reconstruction quality in the case of $R_t = 0.2$ is poorer than the other two cases with $R_t$ values of 0.5 and 0.8. In addition, it is noticed that the true snapshot has some noise superimposed on the main features. However, when we truncate the SVD at lower ranks than the maximum rank of the data matrix, we remove the modes which represent the noise embedded within the data. Consequently, this process leads to the removal of noise from the data, which is another important application of the SVD. This denoising effect is clearly observed in the reconstructed snapshots in the cases of $R_t = 0.5$ and 0.8 across various ranks.

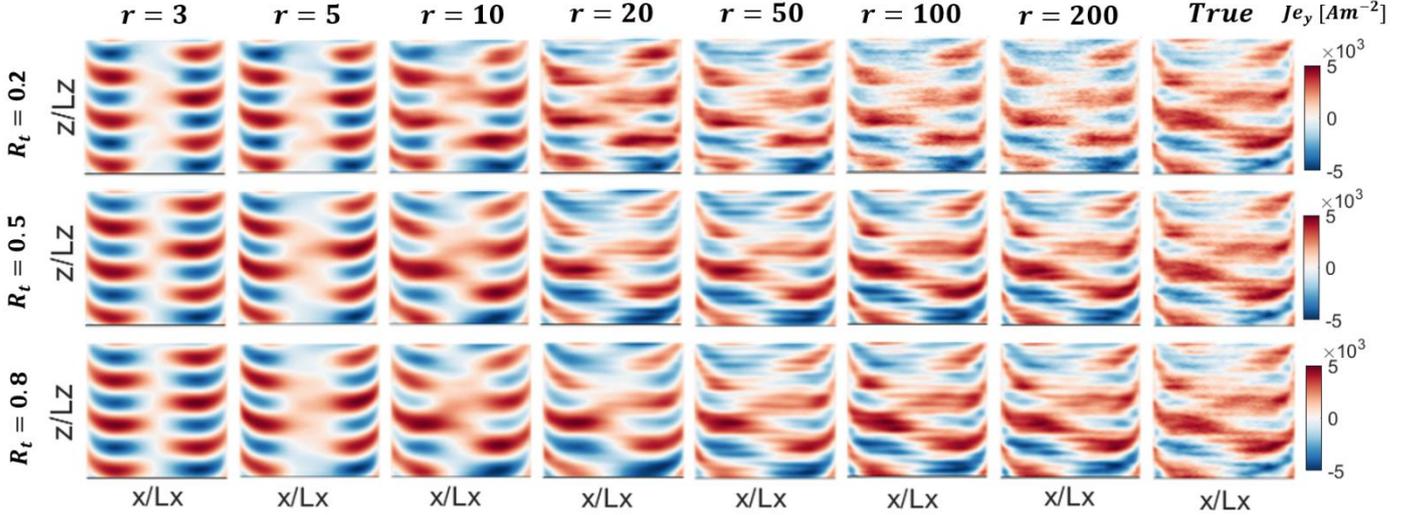

Figure 4: Comparison of the reconstructed 2D snapshots of the axial electron current density ($J_{ey}$) from the SVD using different number of ranks and $R_t$ ratios against the corresponding ground-truth snapshot (last column) from the quasi-2D PIC simulation of the radial-azimuthal baseline test case.

To conclude this subsection, we have shown in Figure 5 a set of SVD spatial modes calculated for the axial electron current density ($J_{ey}$) as well as the azimuthal electric field ($E_z$) data from the baseline radial-azimuthal test case. These modes are among the first 50 SVD modes, which represent distinct patterns within the data. The illustrated modes capture the most prominent structures in the data, which resemble purely azimuthal Electron Cyclotron Drift Instability (ECDI) [19][26][27], radial-azimuthal Modified Two Stream Instability (MTSI) [19][24][28][29], and the transition states between the two instabilities. In any case, we once again emphasize that SVD modes are static and do not inherently capture the evolution of the recovered patterns over time. Whereas, the DMD modes are dynamic, which implies that the DMD is a more suitable approach to extract the spatiotemporal coherent modes as we will demonstrate subsequently.

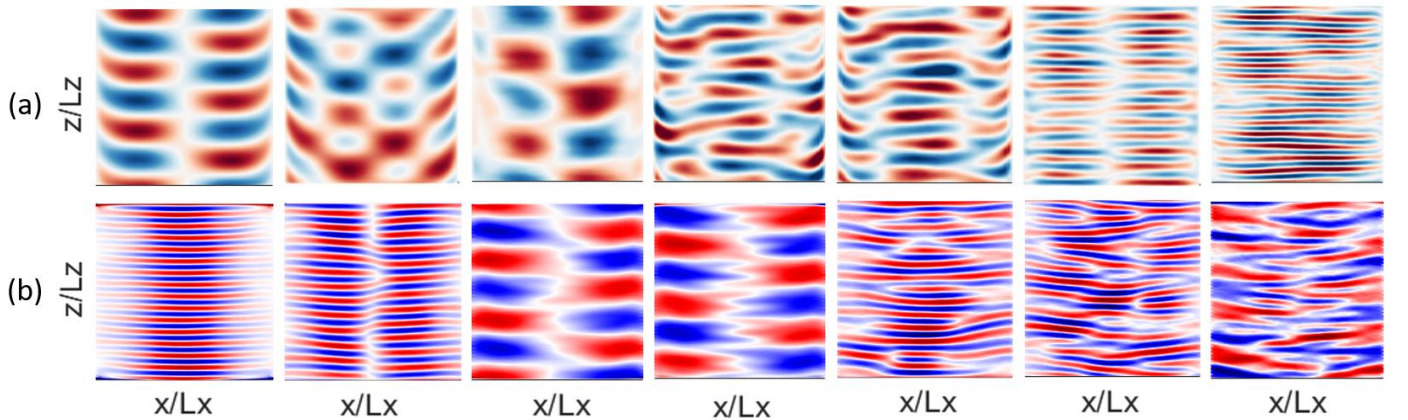

Figure 5: Visualization of seven distinct SVD modes of the ground-truth data from the radial-azimuthal baseline simulation when using $r = 100$ and $R_t = 0.5$ for the SVD. (a) Modes corresponding to the axial electron current density ($J_{ey}$) data, (b) modes corresponding to the azimuthal electric field ($E_z$) data.

## 4.2. Data reconstruction using DMD

In this subsection, we evaluate and compare the ability and performance of the DMD models obtained from the Exact and OPT-DMD methods in reproducing the original dataset by extracting the underlying dominant coherent structures and patterns in the data.

### 4.2.1. Comparison of the OPT-DMD and the Exact DMD methods with different signal durations

To compare the data reconstruction ability of the Exact and OPT-DMD models, we computed the Exact DMD and OPT-DMD of the $J_{ey}$ dataset from the baseline test case with various number of snapshots ($m_t$) corresponding to different lengths of the averaged $J_{ey}$ signal. The respective DMD models were then used to reproduce the data using 75 and 150 DMD modes (or, alternatively, the truncation rank, $r$). The time evolution of the spatially averaged reconstructed and original data is shown in Figure 6. The plots show that only for short time duration ($m_t = 400$) and with relatively large number of modes ($r = 150$), the Exact DMD model yields similar reconstruction capability as the OPT-DMD model. However, over longer durations (larger $m_t$ values), the Exact DMD model immediately reaches a steady value corresponding to the constant mode representing the mean of the data and fails to follow the trend of the ground-truth. Whereas, the OPT-DMD model with either 75 or 150 modes successfully reproduces the overall behavior in the signal over all tested time durations.

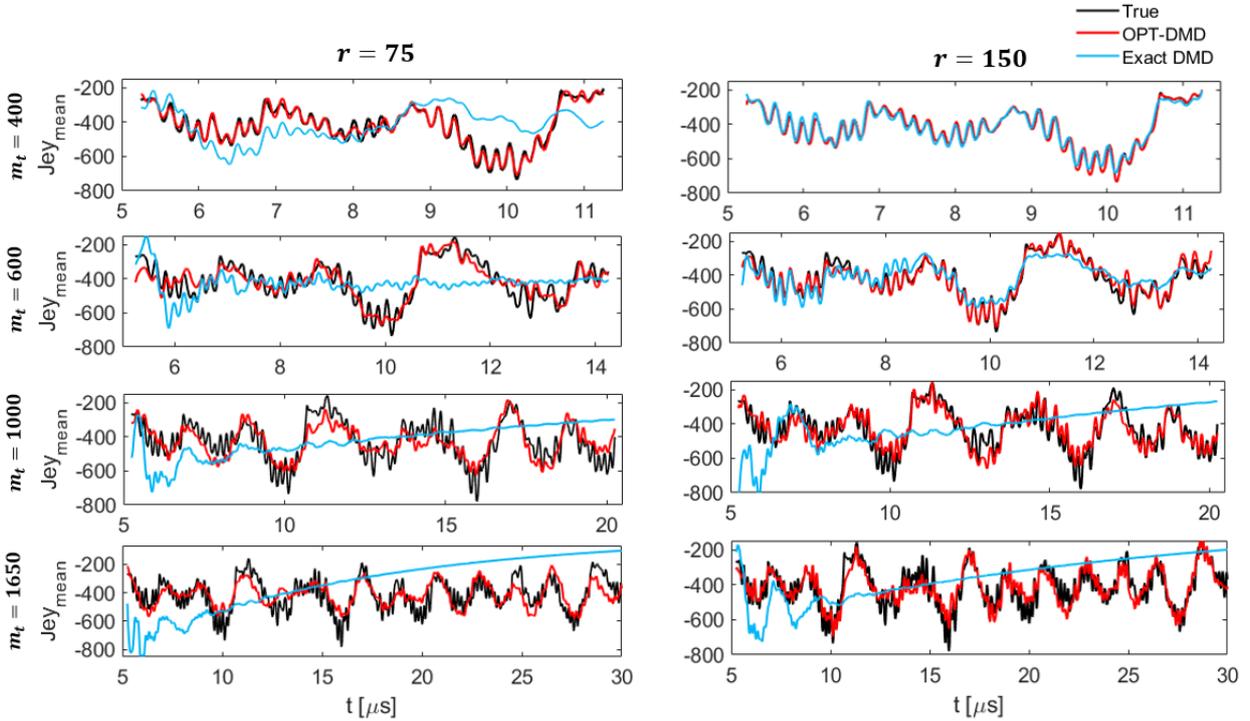

Figure 6: Time evolution of the spatially averaged $J_{ey}$ for the radial-azimuthal test case with the baseline conditions. The rows from top to bottom correspond to different lengths of the training signal (or, equivalently, number of snapshots ($m_t$)). The left and right columns, respectively, correspond to two different DMD ranks of 75 and 150. Ground-truth traces from the PIC simulation are illustrated as solid black lines, solid red lines show the OPT-DMD predicted traces, and the solid blue lines are the predicted traces from the Exact DMD.

The inability of the reconstructed signal from the Exact DMD models to persist over extended times is because the modes' eigenvalues exhibit relatively large negative real values. This results in rapid damping of the oscillatory modes. It is recalled from Eq. 12 that the time dynamics in the DMD method is represented as $exp(\omega_i t)$, hence, it is the real frequency component in this context that denotes growth or damping. The OPT-DMD, nonetheless, stabilizes the model by enabling us to restrict the eigenvalues to lie on the imaginary axis. As a result, the OPT-DMD modes are sustained over arbitrarily long durations without vanishing to zero or growing indefinitely over time. To illustrate this, Figure 7 shows the comparison of the imaginary vs real frequency of the modes obtained with the Exact DMD and the OPT-DMD methods applied to the $J_{ey}$ data of the baseline test case.

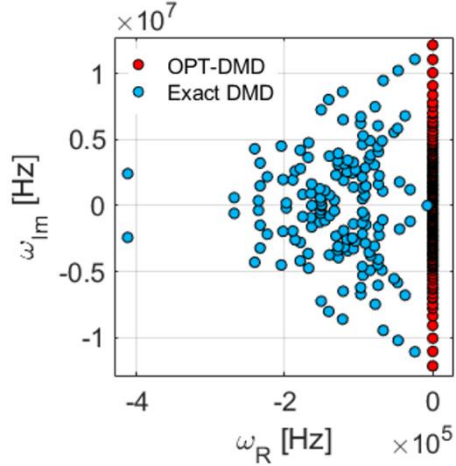

Figure 7: Distribution of the eigenvalues over the complex frequency plane from the Exact DMD and the OPT-DMD algorithms. The eigenvalues correspond to the DMD of the $J_{ey}$ data for the radial-azimuthal baseline test case with the rank of 150 and the training snapshots number of 1650 (Second-column, last-row plot in Figure 6).

The results of this subsection highlight that the OPT-DMD is notably more reliable for data reconstruction and pattern extraction than the Exact DMD. Thus, we will base the remainder of this article on investigations using the OPT-DMD.

**4.2.2. Convergence of OPT-DMD vs truncation rank**

The DMD convergence vs the number of modes (or the truncation rank of the underlying SVD) is an important consideration when approximating complex dynamical systems. The number of modes chosen for the DMD determines the accuracy and fidelity of the resulting approximation. Therefore, as the final assessment before utilizing the OPT-DMD to identify the spatiotemporal modes in our plasma system for practical purposes, we demonstrate the convergence characteristics of the OPT-DMD model as the number of modes increases.

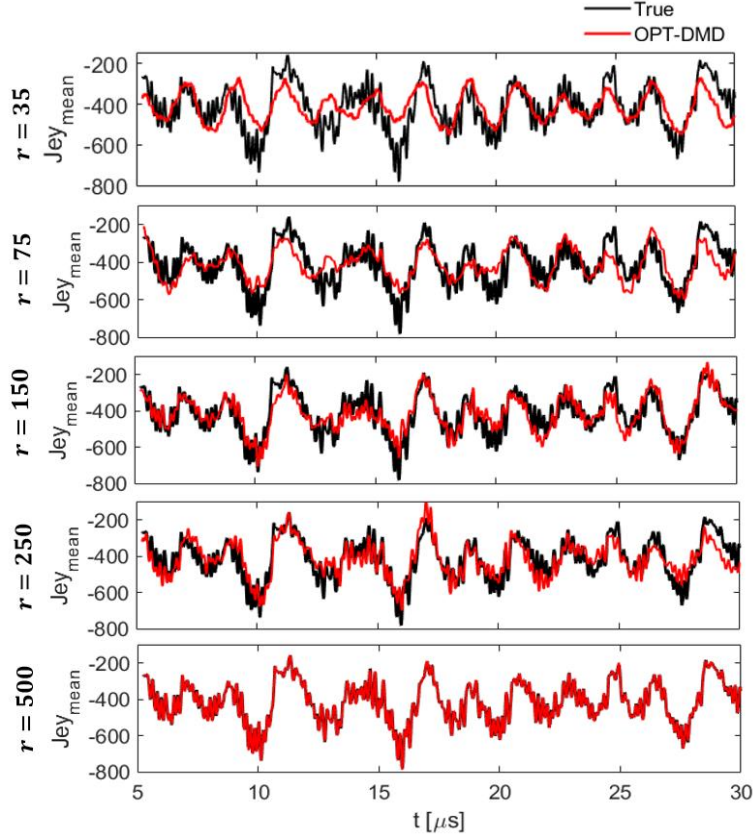

Figure 8: Time evolution of the spatially averaged axial electron current density ($J_{ey}$) for the radial-azimuthal baseline test case from the PIC simulation (solid black line) and the OPT-DMD model (solid red line) with various number of DMD ranks.

Here, we are not concerned with the predictive capability of the model beyond the training dataset as this aspect will be addressed in part II of this paper. Rather, our current focus is on evaluating the loss associated with the reconstruction of the dataset with various number of modes. To this end, we present in Figure 8 a comparison of the time evolution of the spatially averaged $J_{ey}$ data from the baseline test case, reconstructed with different number of DMD modes, against the respective ground-truth data. In addition, samples of reconstructed 2D snapshot with various number of modes corresponding to those shown in Figure 8 are depicted in Figure 9.

From Figure 8 and Figure 9, we can observe that, as the number of modes increases, the DMD algorithm captures finer details present in the data. This allows for a more accurate representation of the system's behavior. Selecting too few modes leads to underfitting, resulting in an oversimplified representation of the system and the loss of critical information. For instance, in the cases with $r = 35$ and $75$, the lower energy modes such the short wavelength azimuthal oscillations superimposed on the main radial-azimuthal patterns are not adequately captured.

It must be noted, however, that too many number of modes can also lead to overfitting, where the model fits the noise in the data rather than the true underlying dynamics. This might cause spurious oscillations or divergence of the models' forecast beyond the training data. Thus, finding the optimal number of modes involves a trade-off between capturing the complexity of the system and avoiding overfitting. The cross-validation by assessing the ability of the DMD model to generalize well to the unseen data is an indication of a proper mode number selection. This is particularly relevant when considering the application of DMD model for forecasting which will be extensively explored in part II.

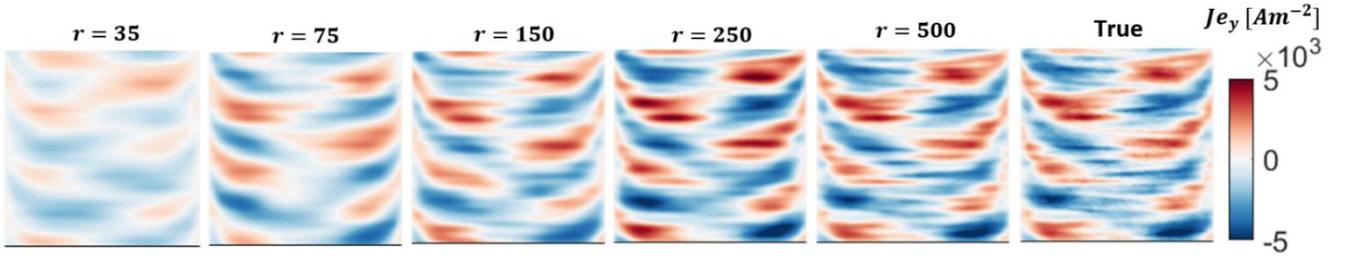

Figure 9: Visualization of a sample reconstructed axial electron current density ($J_{ey}$) snapshot for the baseline test case from the OPT-DMD model with different number of ranks. The subplot denoted as "True" represents the same snapshot from the PIC simulation.

To summarize the conclusions of the above discussions, Figure 10 represents the convergence of the OPT-DMD's average loss (Eq. 15) vs the number of modes used to reconstruct the data. The loss factor continuously decreases with increasing $r$ value, and the rate of decrease in the reconstruction loss also increases for higher number of DMD ranks, especially from about $r = 150$ forward.

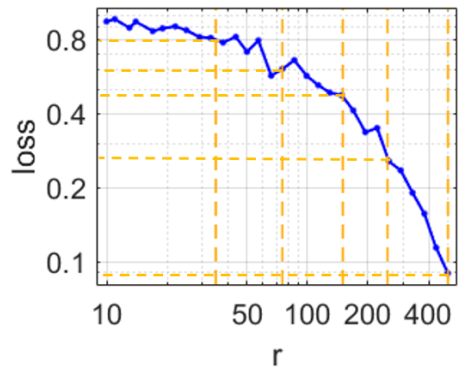

Figure 10: Variation vs rank number ($r$) of the loss factor (Eq. 15) between the ground-truth $J_{ey}$ data from the PIC simulation of the radial-azimuthal baseline test case and the reconstructed $J_{ey}$ data from the OPT-DMD model.

### 4.3. Pattern extraction and coherent modes identification

It was emphasized through the discussions so far that the DMD method identifies the dominant coherent structures (patterns, or modes) in the data and associates each derived mode with linear dynamics, i.e., a complex frequency, with the imaginary part representing the frequency of temporal oscillations and the potential real part representing growth or damping of that mode. The power of the OPT-DMD method was highlighted to partially lie in the fact that the real frequency component of the modes can be minimized and/or eliminated through the optimization process, thus, enabling the OPT-DMD modes to remain fully stable.

In this subsection, we present a novel and practically significant use case of the pattern extraction using OPT-DMD. Indeed, by comparing the discrete spectrum of the DMD modes' frequencies against the temporal FFT of the plasma properties' evolution across various test cases, we demonstrate that the OPT-DMD allows us to derive information on the spatiotemporal characteristics of the involved plasma phenomena such as fluctuations and instabilities in a more accessible and comprehensive manner beyond what is feasible with conventional FFT analyses.

It is noteworthy before delving into the results that the generic spatial vector bases that the DMD method employs to reconstruct and represent a dataset allows this method to circumvent certain limitations of the spatial FFT analysis carried out to determine the wavenumber characteristics of the phenomena. Indeed, the DMD is not limited by the FFT's strict requirement for data periodicity in space as has been pointed out in the literature [30].

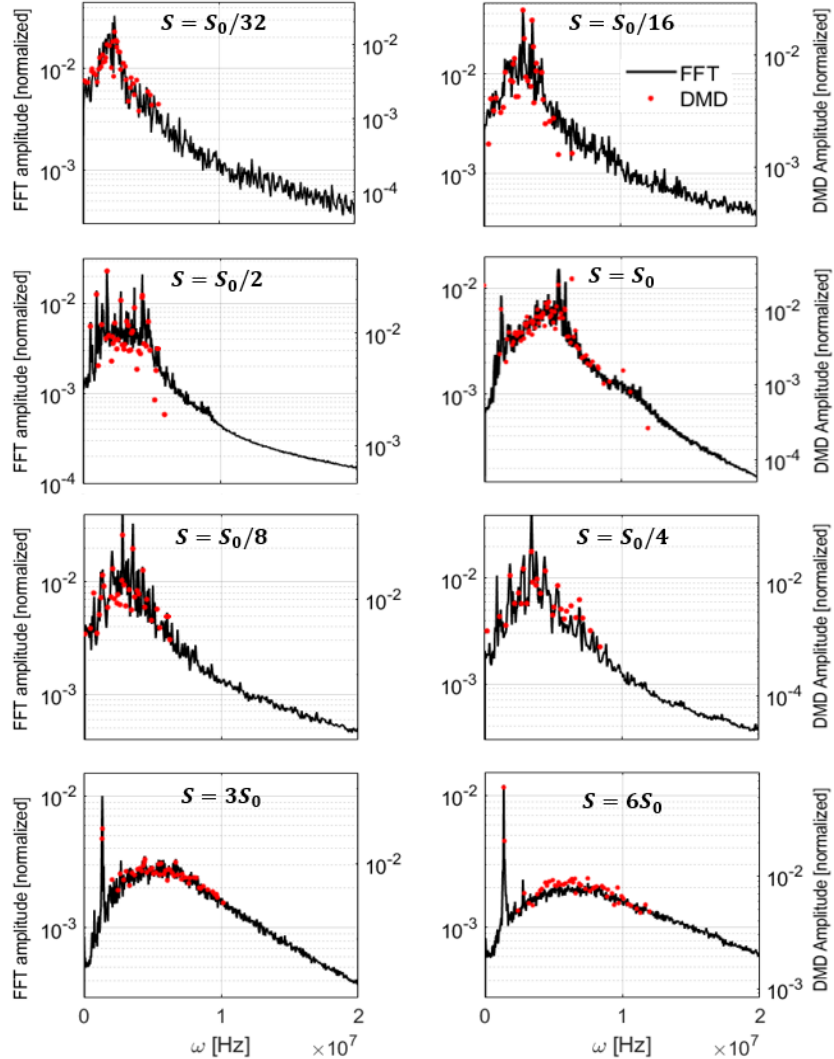

Figure 11: Comparison of the temporal FFT spectra of the azimuthal electric field ($E_z$) for different radial-azimuthal test cases against the distribution of the DMD modes' frequency. The solid black lines represent the FFT of the ground-truth data from the PIC simulations, whereas the red dots correspond to frequency spectra of the $E_z$ DMD modes. Depending on the plasma density across various test cases, the number of DMD ranks used vary between 80 to 150.

In Figure 11, the frequency spectra of the modes obtained by applying OPT-DMD to the $E_z$ data for various test cases are superimposed on the temporal FFT spectra of the same data. It is noticed that, for all cases, the frequency spectra of the OPT-DMD and the FFT are highly consistent. In particular, there are distinct frequency peaks associated with the dominant modes from the OPT-DMD that coincide with the FFT frequency peaks. This superimposition of the frequency spectra allows us to pinpoint the OPT-DMD modes that correspond to dominant and distinct instability/fluctuation modes in the plasma discharge. Following this identification, the frequency information and the spatial structure of the OPT-DMD modes enable a comprehensive, simultaneous, and quantitative spatial-temporal characterization of the instabilities.

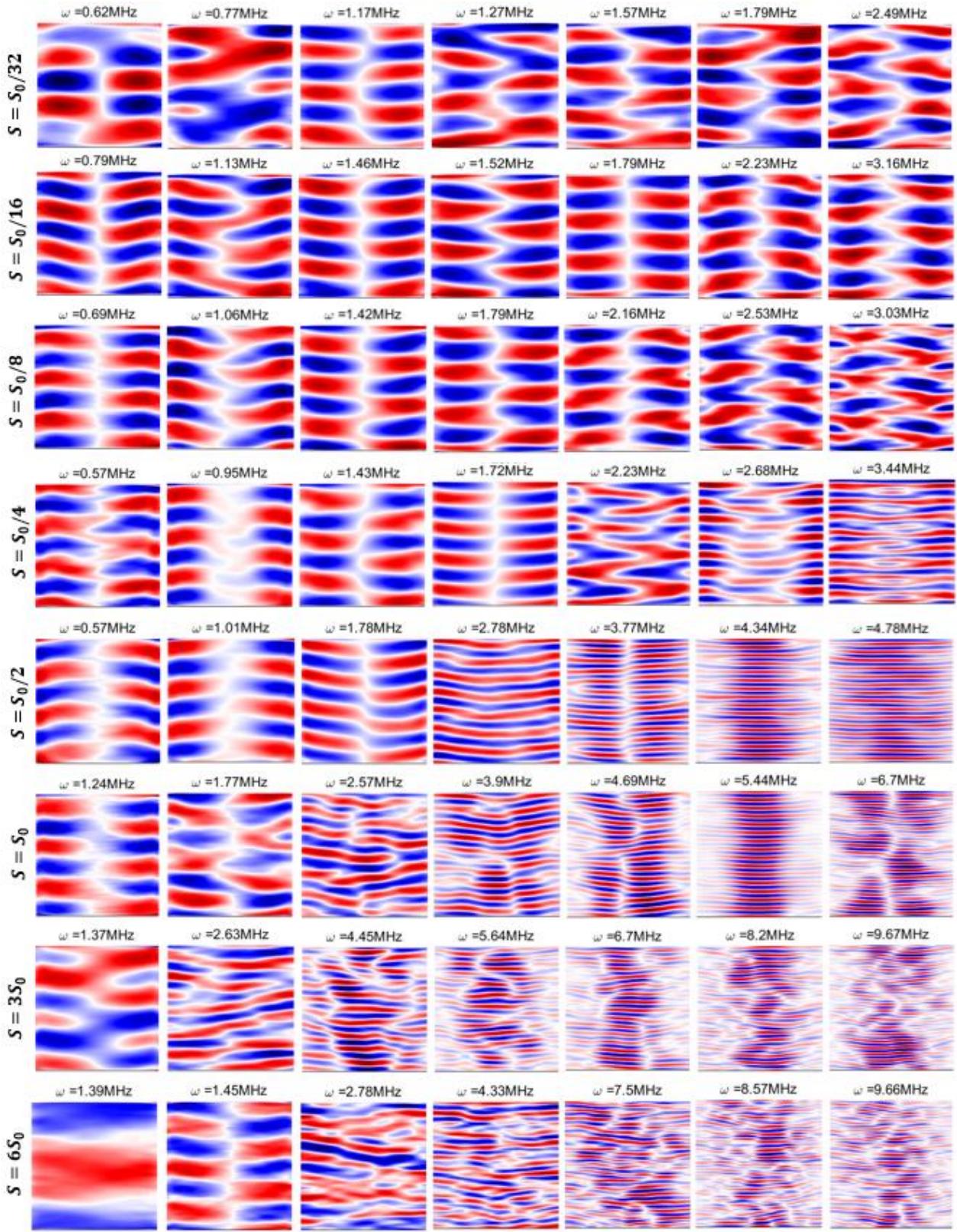

Figure 12: Visualization of the dominant DMD modes of the azimuthal electric field data for various radial-azimuthal test cases that are coincident with the peaks in the temporal FFT plot of each test case.

This is better illustrated in Figure 12, in which, for each radial-azimuthal test case, we have shown the spatial structure of the dominant OPT-DMD modes whose frequencies coincide with the peaks in the FFT plot of each test case. From these 2D mode structures, the wavelength of each specific mode can be readily determined. Plots similar to those shown in Figure 12 can have a remarkable applied significance in that they allow informing experiments by, for instance, providing information on the spatial spacing needed between electrostatic Langmuir probes along each physical direction of the system to experimentally capture a specific fluctuating pattern.

It is also noteworthy that the information presented by the plots in Figure 11 and Figure 12 fully supports the findings of the our previous publication [19], in which we had in part assessed the influence of the plasma number density (value of $S$) on the characteristics of the ECDI and MTSI modes. In that article, we had resorted to spatial and temporal FFT analyses of the PIC simulations' data. These figures, however, provide the same insights in a more visually accessible and easier to comprehend manner. In line with the outcomes of that article [19], it is evident from the plots in Figure 12 that, at low plasma densities ($S$ values from $S_0/32$ to $S_0/8$), the MTSI-like patterns are the only present modes, whereas at high plasma densities ($S$ values of $3S_0$ and $6S_0$), a long-wavelength mode appears that coexists with a spectrum of shorter-wavelength wave modes.

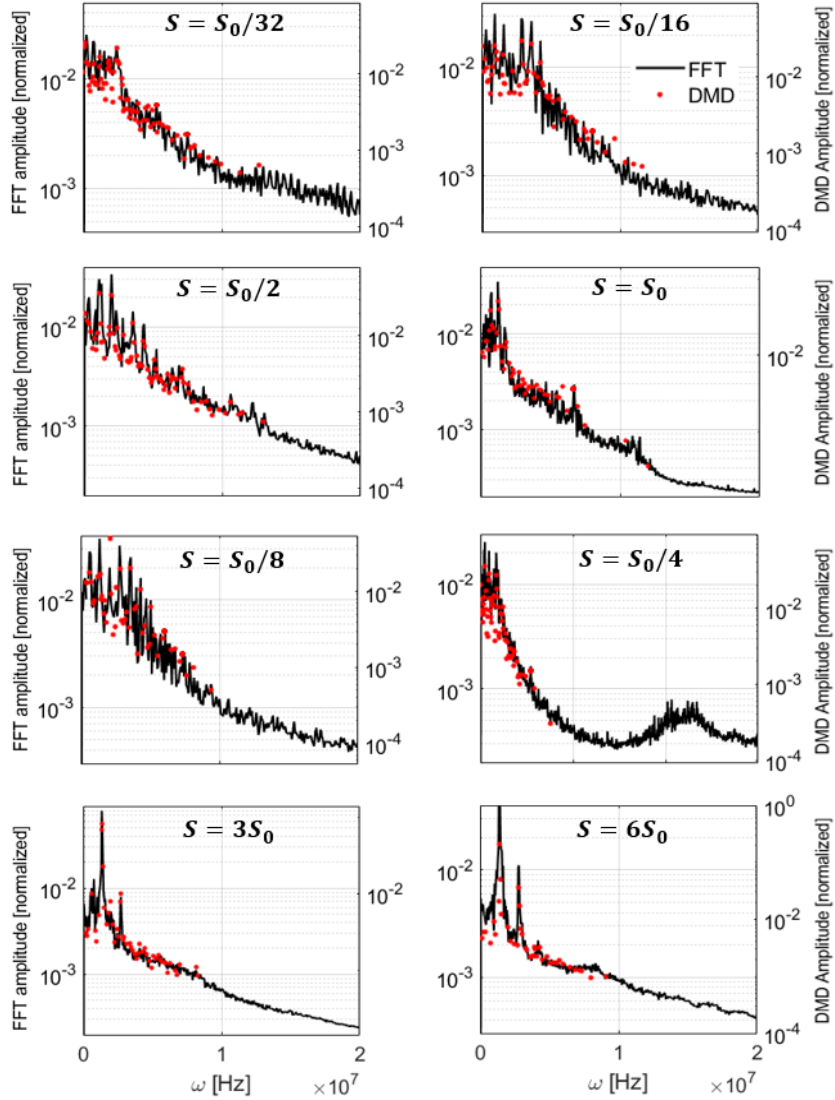

Figure 13: Comparison of the temporal FFT spectra of the axial electron current density ($J_{ey}$) for different radial-azimuthal test cases against the distribution of the DMD modes' frequency. The solid black lines represent the FFT of the ground-truth data from the PIC simulations, whereas the red dots correspond to frequency spectra of the $J_{ey}$ DMD modes. Depending on the plasma density across various test cases, the number of DMD ranks used vary between 80 to 150.

Even though in the demonstrated test cases, the FFT and OPT-DMD may appear as equally well suited for spatiotemporal characterization of observed oscillatory phenomena, in cases where the underlying physics is less known and/or the FFT might become crowded with numerous peaks, the OPT-DMD finds its main advantage.

As one final point, it is observed in plots of Figure 11 that, occasionally, a single frequency peak in the FFT corresponds to several DMD modes. These are manifestations of the DMD rank inflation due to the travelling nature of the waves in our plasma system. The influence of rank inflation on the derived spatiotemporal characteristics of the instability modes is assessed in Section 4.5.

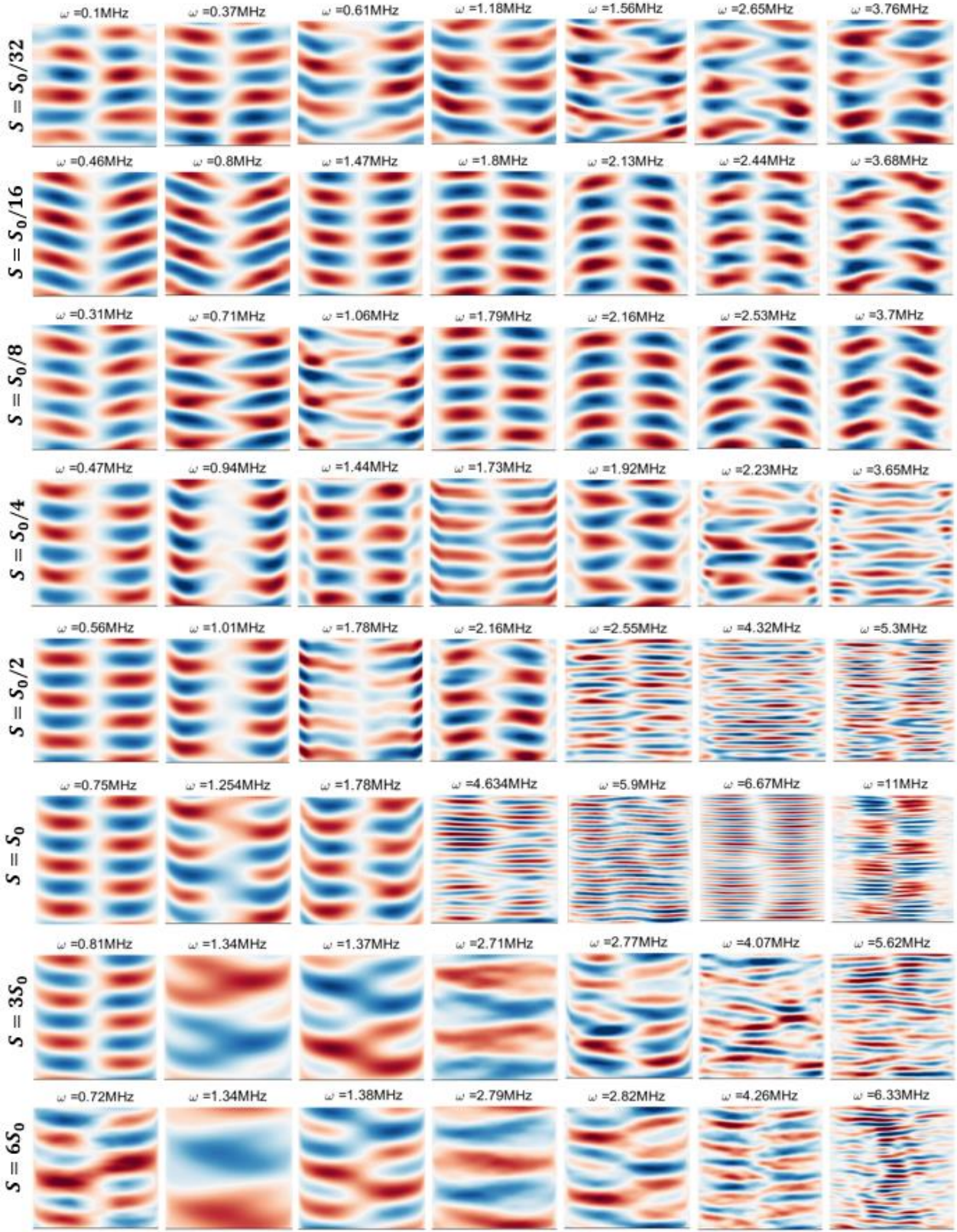

Figure 14: Visualization of the dominant DMD modes of the axial electron current density data for various radial-azimuthal test cases that are coincident with the peaks in the temporal FFT plot of each test case.

The plots in Figure 13 and Figure 14 are provided to further support the arguments presented above by showing similar information but from the application of OPT-DMD to the $J_{ey}$ data of various radial-azimuthal test cases. It is particularly noted from Figure 14 that the dominance of the MTSI-like structures in cases with $S$ values from $S_0/32$ to $S_0/8$ and the appearance of long-wavelength modes for cases with $S$ values of $3S_0$ and $6S_0$, as reported in Ref. [19], can again be clearly discerned.

We would also mention that, toward fully understanding the underlying processes in a system, applying the OPT-DMD to different plasma property data can be advantageous since certain physical information and patterns may be best represented within the evolution data of a specific plasma property.

**4.4. Analysis of the sensitivity of the DMD-identified spatial modes and frequencies to the number of ranks**

In this subsection, we assess the sensitivity to the truncation rank of the identified modes using the OPT-DMD and their spatial structures and frequencies. We performed OPT-DMD on the $J_{ey}$ data of the baseline test case with different number of ranks and compared the characteristics of the most dominant modes corresponding to four distinct peaks in the temporal FFT spectrum of the data.

Figure 15 illustrates the spatial patterns of the modes, their corresponding frequencies, and the discrete DMD spectra with different number of ranks. It is first observed that, for each value of $r$, the DMD spectrum is in agreement with the FFT plot, and that the high-energy modes coinciding with the FFT peaks are represented. Second, the spatial structures of the modes are similar when DMD is calculated with different number of ranks. Third and last, the frequencies associated with each mode show a variation of around 1-6 % across all cases.

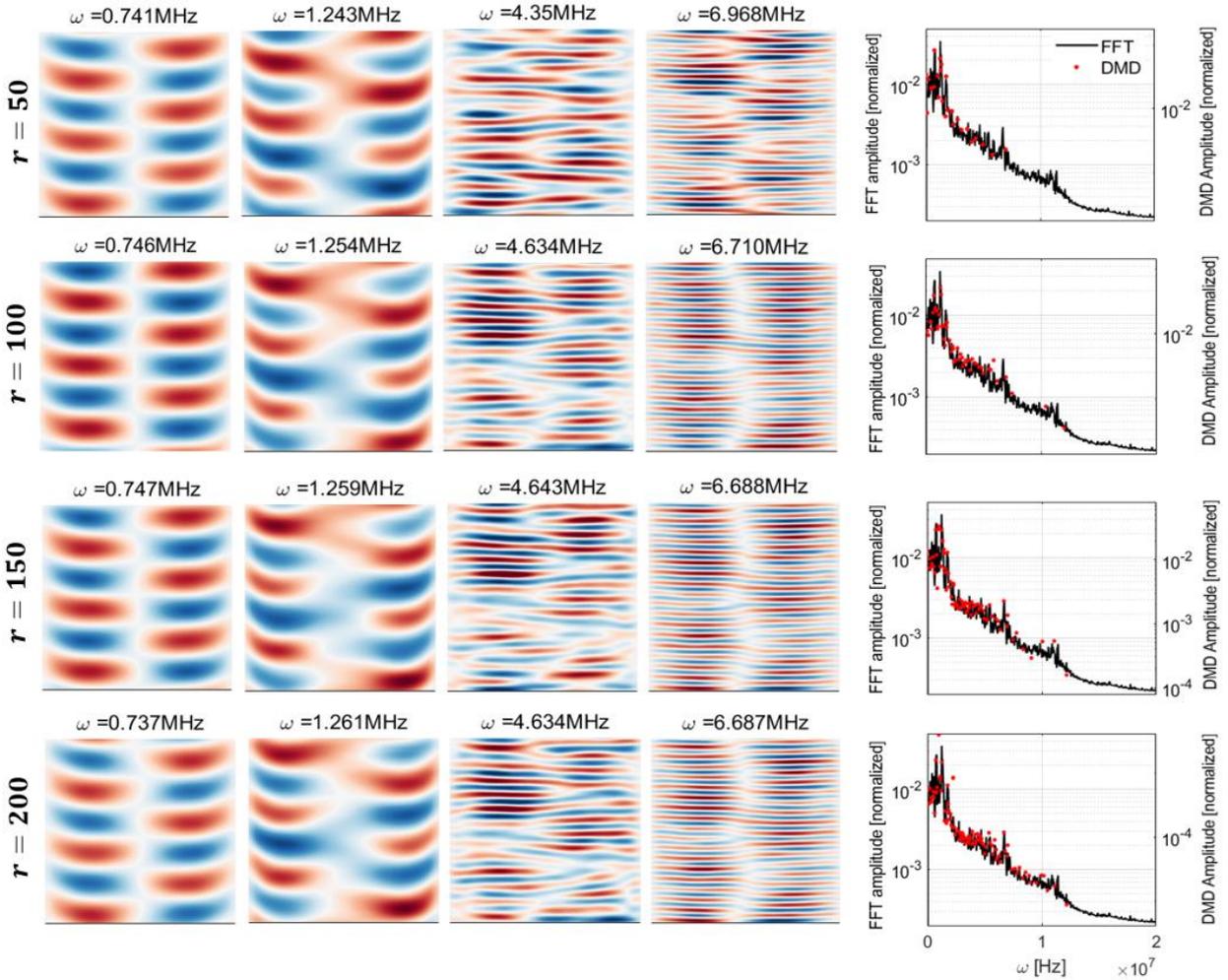

Figure 15: Effect of the number of ranks on the DMD frequency spectra against the temporal FFT and on the shape of the sample dominant frequency modes. The test case corresponds to the baseline conditions. The FFT and the DMD are applied to the $J_{ey}$ data.

Based on these observations, it can be concluded that the spatiotemporal characteristics of the modes exhibit a very low sensitivity across different choices of the number of ranks for the DMD model. This implies that the most significant features are captured even when the number of retained modes varies. The consistency of the modes' characteristics regardless of the chosen rank underscores the robustness of the OPT-DMD method for the task of dominant patterns extraction and modes identification.

### 4.5. Analysis of the effect of rank inflation on the DMD-identified spatial modes

It is well known that a fundamental limitation of SVD-based approaches, such as the DMD, is their ineffective treatment of translational/rotational invariances in the data. The reason is that the SVD relies on the correlation between rows and columns of the data matrix, and any misalignment in the data can lead to an overestimation of the rank of the data due to the inclusion of extra variations rooting in translation and/or rotation. This has an implication on the effectiveness of dimensionality reduction we hope to achieve with SVD-based methods for a data that features fundamentally low-rank objects shifted through space. In such cases, the SVD struggles to accommodate the shifts or translational invariances in the patterns, and instead represent the object with an increased number of ranks, leading to an artificial rank inflation. A relevant example of this in the plasma applications is the existence of travelling waves.

Looking at the DMD spectra in Figure 11 and Figure 13, we notice that, around the FFT peaks, the DMD spectra often contain multiple modes. These instances correspond to dominant travelling waves, to represent which the DMD requires several modes.

Here, we demonstrate the rank inflation by showing in Figure 16 the spatial patterns of a group of modes that represent the first four dominant DMD modes of the $J_{ey}$ data for the baseline test case calculated using 200 ranks.

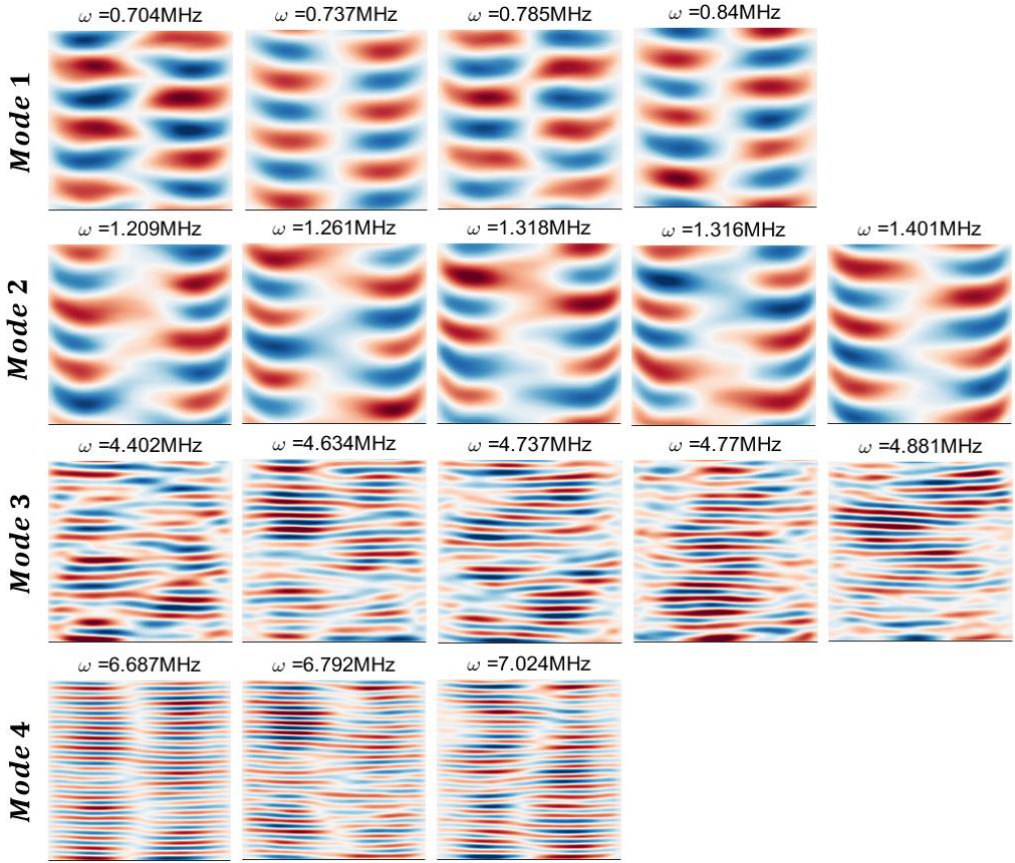

Figure 16: Demonstration of the effect of DMD rank inflation, and the similarity of the shape of the DMD modes at close frequencies. The distinct $J_{ey}$ DMD mode numbers 1 to 4 correspond to those on the bottom row of Figure 15.

Each row in Figure 16 illustrates the set of modes with close frequencies to a certain peak in the FFT spectrum (Figure 15, last row), hence, representing a single wave. It is evident that each group of modes share similar spatial characteristics except that they are shifted along the azimuthal direction. Therefore, although the rank inflation due to presence of travelling waves artificially increases the dimensionality of the system's dynamics as captured by the OPT-DMD method, it does not raise any concern regarding modes identification since selecting any mode from a "mode cluster" equivalently provides similar information about the spatial structure of the wave.

**Section 5: Conclusions**

In this article, we provided a gentle applied introduction to the concepts and mathematical foundations of the SVD/POD and DMD, underlined the relationship between these data-driven methods, and presented the application of the SVD/POD and DMD toward dimensionality reduction of high-dimensional plasma simulations' data as well as pattern extraction and modes identification.

Regarding the SVD/POD, we demonstrated how the method enables understanding the nature of the data in terms of low-rank (low-dimensional) representability and can be used for data compression and de-noising. We also showed the variation in the capability of the SVD to represent and reconstruct unseen (test) data based on the reduced coordinates it has learned from the ground-truth data at different training-to-test ratios.

Concerning the DMD, we compared the performance of the Exact (basic) and the OPT-DMD variants of the method toward low-rank dynamics representation of the ground-truth data. We observed that OPT-DMD provides a much more reliable means to carry out this task for multiple values of training dataset size and number of DMD ranks. This was highlighted to root in the fact that the OPT-DMD enables constraining the eigenvalues (frequencies) associated with the modes to near the imaginary axis, thus, allowing the derived DMD modes to remain stable over arbitrary time durations and do not tend to zero or diverge to infinity as is often the case with the Exact DMD method when applied to noisy data of real-world systems. These results emphasized the noise robustness of the OPT-DMD which makes this variant well suited for applications to the plasma systems' data.

For the OPT-DMD variant, we evaluated the convergence characteristics of the algorithm against the number of ranks. It was pointed out that, even though increasing the truncation rank improves consistently the similarity between the OPT-DMD-reconstructed and the ground-truth data, a very high rank number can lead to overfitting the training data by forcing the algorithm to learn the noise as well. This may in turn affect the forecasting accuracy of an OPT-DMD model as it will be discussed in detail in the article's part II. However, too few number of ranks leads to the loss of important information in the data, and it is for these reasons that the selection of the optimal number of DMD ranks comes as a trade-off between the accuracy of the training data reconstruction and the prediction of the test data.

Next, we presented a novel application of the OPT-DMD method to complement and/or augment the spectral analyses conventionally carried out using FFT technique. We demonstrated across various test cases the consistency between the frequency spectra of the modes from the OPT-DMD and the FFT of plasma data (azimuthal electric field and axial electron current density), illustrating how the OPT-DMD enables a more visually accessible, simultaneous characterization of the spatial-temporal properties of the plasma instabilities and fluctuations. The important benefits of the use of OPT-DMD for spectral analyses of data in terms of informing the experimental characterizations of plasma oscillations were underlined.

We analyzed the sensitivity to the number of DMD ranks of the spatial structure and frequency of the identified OPT-DMD modes, demonstrating that the modes' characteristics from the OPT-DMD is highly robust with respect to the truncation rank. Moreover, we assessed the influence of rank inflation on the identification of physically meaningful dominant coherent structures in the plasma data using OPT-DMD. We showed that different DMD modes with frequencies around any specific peak in the temporal FFT plot of the data feature very similar spatial distributions, meaning that the rank inflation does not affect the reliability of the OPT-DMD to be used for the inference of the spatiotemporal characteristics of the instabilities.

**Acknowledgments**:

The present research is carried out within the framework of the project "Advanced Space Propulsion for Innovative Realization of space Exploration (ASPIRE)". ASPIRE has received funding from the European Union's Horizon 2020 Research and Innovation Programme under the Grant Agreement No. 101004366. The views expressed herein can in no way be taken as to reflect an official opinion of the Commission of the European Union.

MR, FF, and AK gratefully acknowledge the computational resources and support provided by the Imperial College Research Computing Service (http://doi.org/10.14469/hpc/2232).

**Data Availability Statement**:

The simulation data that support the findings of this study are available from the corresponding author upon reasonable request.

**References**:

[1] Schmid PJ, "Dynamic mode decomposition of numerical and experimental data", *J. Fluid Mech.* **656**.1 (2010)

[2] Rowley CW, Mezic I, Bagheri S, Schlatter P, Henningson D, "Spectral analysis of nonlinear flows", *J. Fluid Mech.* **641** (2009)

[3] Mezic I, "Analysis of fluid flows via spectral properties of the Koopman operator", *Annu. Rev. Fluid Mech.* **45**:357-378 (2013)

[4] Brunton SL, Noack BR, Koumoutsakos P, "Machine Learning for Fluid Mechanics", *Annu. Rev. Fluid Mech.* **52** (2020)

[5] Askham T, Kutz JN, "Variable projection methods for an optimized dynamic mode decomposition", *SIAM Journal on Applied Dynamical Systems* **17**:1, 380-416 (2018)

[6] Sashidhar D, Kutz JN, "Bagging, optimized dynamic mode decomposition for robust, stable forecasting with spatial and temporal uncertainty quantification", *Phil. Trans. R. Soc. A* **380**:20210199 (2022)

[7] Sasaki M, Kawachi Y, Dendy RO, Arakawa H, Kasuya N et al, "Using dynamical mode decomposition to extract the limit cycle dynamics of modulated turbulence in a plasma simulation", *Plasma Phys. Control. Fusion* **61** 112001 (2019)

[8] Nayak I, Teixeira FL, "Dynamic Mode Decomposition for Prediction of Kinetic Plasma Behavior", Proceedings of 2020 International Applied Computational Electromagnetics Society Symposium (ACES), Monterey, CA, USA, pp. 1-2 (2020)

[9] Nayak I, Kumar M, Teixeira FL, "Detection and prediction of equilibrium states in kinetic plasma simulations via mode tracking using reduced-order dynamic mode decomposition", *Journal of Computational Physics* **447** 110671 (2021)

[10] Taylor R, Kutz JN, Morgan K, Nelson BA, "Dynamic mode decomposition for plasma diagnostics and validation", *Rev Sci Instrum* **89**, 053501 (2018)

[11] Kaptanoglu AA, Morgan KD, Hansen CJ, Brunton SL, "Characterizing Magnetized Plasmas with Dynamic Mode Decomposition", *Physics of Plasmas* **27**, 032108 (2020)

[12] Brunton SL, Kutz JN, "Data-Driven Science and Engineering: Machine Learning, Dynamical Systems and Control", Second edition, Cambridge University Press, ISBN: 1009098489, 2022.

[13] Kaganovich ID, Smolyakov A, Raitses Y, et al., "Physics of E × B discharges relevant to plasma propulsion and similar technologies", Phys. Plasmas 27, 120601 (2020)

[14] Tu JH, Rowley CW, Luchtenburg DM, Brunton SL, Kutz JN, "On dynamic mode decomposition: Theory and applications". *Journal of Computational Dynamics*, **1**(2):391–421 (2014)

[15] Kutz JN, Fu X, Brunton SL, "Multiresolution Dynamic Mode Decomposition", *SIAM Journal on Applied Dynamical Systems* **15**:2, 713-735 (2016)

[16] Le Clainchey S, Vegay JM, "Higher Order Dynamic Mode Decomposition", SIAM *Journal on Applied Dynamical Systems* 16:2, 882-925 (2017)

[17] Andreuzzi F, Demo N, Rozza G, "A dynamic mode decomposition extension for the forecasting of parametric dynamical systems", arXiv:2110.09155v1 (2021)

[18] Askham T, duqbo/optdmd: optdmdv1.0.0, https://doi.org/10.5281/zenodo.439385 (2017)

[19] Reza M, Faraji F, Knoll A, "Parametric investigation of azimuthal instabilities and electron transport in a radial-azimuthal E×B plasma configuration", *Journal of Applied Physics* **133**, 123301 (2023)

[20] Reza M, Faraji F, Knoll A, "Concept of the generalized reduced-order particle-in-cell scheme and verification in an axial-azimuthal Hall thruster configuration", *J. Phys. D: Appl. Phys.* **56** 175201 (2023)

[21] Faraji F, Reza M, Knoll A, "Enhancing one-dimensional particle-in-cell simulations to self-consistently resolve instability-induced electron transport in Hall thrusters". *J. Appl. Phys.* **131**, 193302 (2022)

[22] Reza M, Faraji F, Knoll A, "Resolving multi-dimensional plasma phenomena in Hall thrusters using the reduced-order particle-in-cell scheme", *J Electr Propuls* **1**, 19 (2022)

[23] Faraji F, Reza M, Knoll A, "Verification of the generalized reduced-order particle-in-cell scheme in a radial-azimuthal E×B plasma configuration", *AIP Advances* **13**, 025315 (2023)

[24] Villafana W, Petronio F, Denig AC, Jimenez MJ, et al., "2D radial-azimuthal particle-in-cell benchmark for E×B discharges", *Plasma Sources Sci. Technol.* **30** 075002 (2021)

[25] Gavish M, Donoho DL, "The Optimal Hard Threshold for Singular Values is 4/√3", *in IEEE Transactions on Information Theory*, vol. 60, no. 8, pp. 5040-5053, (2014)

[26] Boeuf JP, "Tutorial: Physics and modeling of Hall thrusters", *Journal of Applied Physics* **121** 011101 (2017)

[27] Janhunen S, Smolyakov A, Chapurin O, Sydorenko D, Kaganovich I, Raitses Y, "Nonlinear structures and anomalous transport in partially magnetized E×B plasmas," *Phys. Plasmas* **25**, 011608 (2018)

[28] Janhunen S, Smolyakov A, Sydorenko D, Jimenez M, Kaganovich I, Raitses Y, "Evolution of the electron cyclotron drift instability in two dimensions," *Phys. Plasmas* **25**, 082308 (2018)

[29] Petronio F, Tavant A, Charoy T, Alvarez-Laguna A, Bourdon A, Chabert P, "Conditions of appearance and dynamics of the Modified Two-Stream Instability in E×B discharges", *Phys. Plasmas* **28**, 043504 (2021)

[30] Petronio F, "Plasma instabilities in Hall Thrusters: a theoretical and numerical study", PhD dissertation, Paris Polytechnic Institute, NNT: 2023IPPAX030 (2023)